# Use of synthetic data for training dose estimation neural networks in CT dosimetry

**Marie-Luise Kuhlmann (corresponding author)**

Dosimetry for Radiation Therapy and Diagnostic Radiology,

Physikalisch-Technische Bundesanstalt (PTB), Braunschweig, 38116, Germany

Technische Universität Dortmund, Dortmund, 44227, Germany

**Jörg Martin**

Machine Learning and Uncertainty,

Physikalisch-Technische Bundesanstalt (PTB), Braunschweig, 38116, Germany

**Stefan Pojtinger**

Dosimetry for Radiation Therapy and Diagnostic Radiology,

Physikalisch-Technische Bundesanstalt (PTB), Braunschweig, 38116, Germany

**Running title:** Synthetic data use for dose estimating NNs

*Corresponding author*: Marie-Luise Kuhlmann

                              Mailing address: *Bundesallee 100, 38116 Braunschweig, Germany*

                              Email address:    *marie-luise.kuhlmann@ptb.de*

# Abstract

**Objective**: Personalized computed tomography (CT) dosimetry has great potential in assessing patient-specific radiation exposure, supporting risk assessment, and optimizing clinical protocols. The aim of this study is to evaluate the potential of synthetic anatomical data for improving machine learning-based personalized computed tomography (CT) dosimetry. Specifically, it is investigated whether the combination of synthetic human body geometries with real patient data can improve model accuracy and generalization for CT organ dose estimation while maintaining the uncertainty requirements outlined in IAEA TRS-457.

**Approach**: Deep learning models for organ dose prediction are trained using datasets with varying proportions of real and synthetic data. Synthetic datasets are generated from computational human phantoms with controlled distributions of organ volumes and body shapes to ensure balanced statistical coverage. A dedicated model uncertainty evaluation method is implemented to quantify prediction reliability and verify compliance with TRS-457 accuracy limits. Model performance and uncertainty are compared across different training data compositions, including a model trained solely on real patient data. As baseline validated Monte Carlo simulation is used.

**Main results**: Models trained solely on synthetic data show limited predictive accuracy, particularly for small or peripheral organs. Incorporating as little as 10 % real patient data significantly improves both statistical accuracy and uncertainty estimates, achieving a performance comparable to that of real-only models. The hybrid training approach improves robustness across different anatomies while maintaining TRS-457-compliant uncertainty levels ($k = 2$ uncertainty < 20% for adults). The results indicate that the combination of real and synthetic data in combination with a systematic uncertainty assessment supports the development of CT dosimetry models and at the same time reduces the amount of real data required.

# 1. INTRODUCTION AND PURPOSE

A well-structured patient dose database for computed tomography (CT) scans can serve as a resource for researchers investigating the effects of low-dose exposure, as an aid in the optimization of clinical scanning protocols, and as a tool for deriving diagnostic reference levels. Assessing dose values requires user-friendly methods that operate on the small time scales essential for successful integration into the clinical workflow. The IAEA states in TRS-457 that a 20 % ($k = 2$) uncertainty for summed organ doses is sufficient for radiological examinations in adults. Since children and pregnant patients are more sensitive to radiation, a dosimetric uncertainty of 7 % ($k = 2$) is considered sufficient for assessing their risk. For comparative assessment, a 7 % ($k = 2$) uncertainty is considered acceptable [1].

One method for implementing fast and convenient dose calculation is through machine learning. The use of machine learning algorithms in medical applications is becoming increasingly common, and several publications have demonstrated the suitability of neural networks for dose estimation tasks (e.g., [2], [3], [4]). In each of these publications, the networks examined were trained on real patient data, and the distribution of the patient data with respect to sex and age are described. Meier *et al* additionally present the effective diameter range of the patients [2]. Other parameters — such as patient height, organ volumes, and adipose tissue — are not addressed. The distributions of these and other parameters are difficult to control and are constrained by data availability.

The distribution of training data is a major factor influencing a network's generalization capability. The underrepresentation of rare cases can lead to unwanted biases [5]. To address this problem, the use of synthetic training data is common in medical applications such as lesion classification [6] and image segmentation [7]. The work by Russ *et al* [8] showcased the feasibility of using CycleGANs to generate realistic CT images from XCAT phantoms, achieving high anatomical accuracy and demonstrating improved segmentation performance when synthetic and real data were combined. Another example is the 2024 study by Khosravi *et al*, which showed that adding synthetic chest X-rays to the CheXpert and MIMIC-CXR datasets improved performance, with gains saturating around a 10:1 synthetic-to-real ratio [9].

The use of computationally generated human body geometries such as XCAT [10] enables the creation of a diverse data set with controllable statistical distributions. This allows a balanced distribution to be presented to the network, thus avoiding biased training.

Although synthetic anatomical models and data augmentation have demonstrated potential in machine learning for diagnostic purposes, their use in personalised CT dosimetry is not well established.

The aim of the work presented here is to provide insight into the potential offered by synthetic data in machine learning-based personalized dosimetry and to investigate the value of combining real and synthetic datasets in order to improve generalization and reduce reliance on real patient data. Specifically, it is analysed how synthetic human geometries can support, with respect to the uncertainties mentioned in TRS-457, the development of machine learning models evaluated on real patient data. The focus was on using a set of geometries with well-defined and controlled distributions of organ volumes and body shapes to train deep learning models for CT organ dose estimation. To investigate the usability of the synthetic data, the model performance was evaluated over different training data compositions, including different ratios of real and synthetic patient data. The performance of these models was compared to a baseline model trained exclusively on real patient data.

## 2. MATERIAL AND METHODS

### 2.1. Human geometry dataset

*2.1.1. Synthetic patient geometry dataset*

To generate the phantom data, the commercial software XCAT Version 2 [10] was used. As its input, the software received parameters defining organ volume, fat layer thickness, and body height. The numerical values used to generate the synthetic phantoms were sampled from uniform distributions. The mean value of the organ volume distributions agreed with those reported in ICRP Publication 145 [11]. The range of the assumed uniform distributions was set to ±10 % relative to the mean. To represent variations in body types, the fat layer thicknesses around the abdomen, thorax, and extremities were adjusted using sampled scaling factors in the range of [0.90, 1.1]. 100 male and 100 female adult phantoms were generated, with body heights ranging from 1.50 m to 2.10 m. For each parameter 200 values were sampled, and random combinations of these values were used to generate the final phantom set of 200 phantoms. For the fat layer thickness, only one scaling factor was used for all the fat layers located at different positions within a single phantom. The dataset evaluation is presented in [12]. The organ volume distribution for the different datasets is shown in Figure 7.

*2.1.2. Real patient CT scan dataset*

The synthetic dataset was supplemented by 116 whole-body CT scans provided by the Städtisches Klinikum Braunschweig. All scans were deidentified to ensure no personal data was present in the files. The dataset included 54 female and 62 male adult patients. The scan resolution was 1 mm x 1 mm x 1 mm. The CT scans were segmented into 104 organs and tissues using the TotalSegmentator [13] software. For further use, several tissues and organs were combined (see Table 2).

Both datasets, real and synthetic, initially covered the entire body. For this study, however, only the thoracic region was considered. To extract the thoracic region, an automated region-of-interest (ROI) extraction was applied based on anatomical landmarks, selecting slices ranging from the cervical vertebrae to the adrenal gland. The ROI boundaries were then extended by 30 image slices on each side.

## 2.2. Radiation transport simulation

The ground truth for the training of the neural network was generated using radiation transport with the Monte Carlo simulation software EGSnrc [14]. For the simulation of the CT radiation field, a recently implemented and validated particle source [15] was utilized. The particle source can be adapted to a real CT scanner using measured data as the input parameters for the source. The input parameters were the collimation, the distance between the source and the gantry centre, and the spectrum and aluminium equivalent representation of the bowtie filter.

Table 1: Monte Carlo transport parameters and characteristics as defined in [16].

| Parameter | Description |
|---|---|
| Code | EGSnrc release version 2021[14] |
| User code | Cavity |
| Cross-sections | mcdf-xcom [17] |
| Transport parameters | Electron cutoff 0.512 MeV (includes rest energy) |
|  | Photon cutoff 0.001 MeV |
| Scored quantities | Dose per particle for each voxel |
|  | Saved in a .3ddose file |
| Variance reduction | None |
| Statistical uncertainty | Inside the body: 0.5 – 2 % |
| Postprocessing | Conversion to Nifti-format |

For the simulations performed for this work, the source was adapted to simulate an Optima CT 660 (GE Healthcare). The transport parameters are shown in Table 1.

The phantom data and the segmented real CT scans were converted to the EGSnrc geometry format egsphant. The used materials and densities are presented in Table 2.

Table 2: Material definition for radiation transport simulation.

| Material | Density correction file | Density in g/cm$^3$ | Electron density in m$^{-3} \cdot 10^{26}$ |
|---|---|---|---|
| Air | air_dry_nearsealevel | 0.0012 | 3.3 |
| Muscle | muscle_skeletal_icrp | 1.04 | 3480 |

| Lung | lung_inflated_icru_1986 | 0.27 | 862 |
| Spine Bone | bone_cortical_icrp | 1.85 | 5950 |
| Rib Bone | bone_compact_icru | 1.85 | 5950 |
| Fat | adiposetissue_icrp | 0.92 | 3180 |
| Blood | blood_icrp | 1.06 | 3510 |
| Heart | heart_blood-filled_icru_1986 | 1.06 | 3510 |
| Kidney | kidney_icru_1986 | 1.05 | 3480 |
| Liver | liver_icru_1986 | 1.06 | 3510 |
| Pancreas | pancreas_icru_1986 | 1.04 | 3460 |
| Intestine | gitract_intestine_icru_1986 | 1.03 | 3420 |
| Scull | bone_compact_icru | 1.85 | 5950 |
| Brain | brain_icrp | 1.03 | 3460 |
| Spleen | spleen_icru_1986 | 1.06 | 3510 |

## 2.3. Neural network training

### 2.3.1. Network architecture

A three-dimensional U-Net architecture [18] based on the PyTorch Lightning library was used for the dose prediction network. The architecture followed the U-Net encoder-decoder principle but was extended to allow for uncertainty quantification.

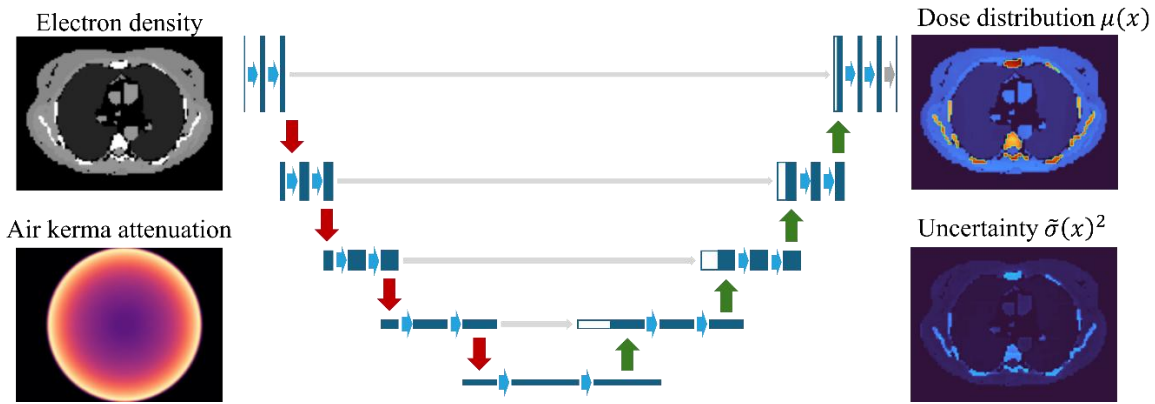

Figure 1 sketches the structure of the neural network. The encoder path consisted of four convolutional blocks. Each convolutional block consisted of two successive 3D convolutions, with each of these followed by batch normalization and a ReLU activation. Downsampling was realized by max-pooling layers with a kernel size of 2×2×2 to reduce spatial dimensions while increasing feature depth. In the decoder path, transposed convolutions ("deconvolutions") with a kernel size of 2×2×2 were used to restore spatial resolution.

The final output layer has two channels: The first channel corresponds to the dose map $\mu(x)$, while the second channel $\tilde{\sigma}(x)^2$ is used to compute logarithmic variance of the dose values $\log \sigma(x) = \log(\tilde{\sigma}(x)^2 + 5.5\% \cdot \mu(x))$, cf. (5) below.

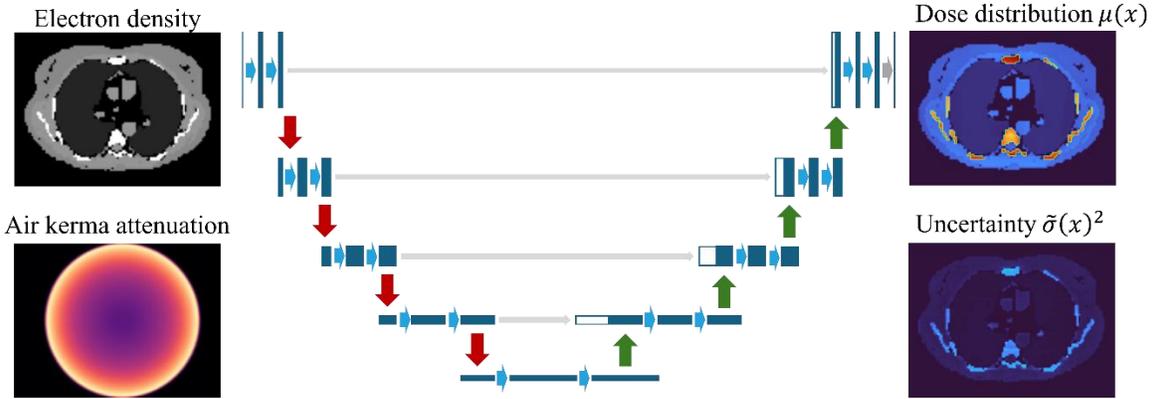

Figure 1. Applied U-Net architecture with a depth of four and illustration of the two inputs electron density and air kerma attenuation map and output features dose distribution map $\mu(x)$ and variance of the output $\tilde{\sigma}(x)^2$.

### 2.3.2. Training data

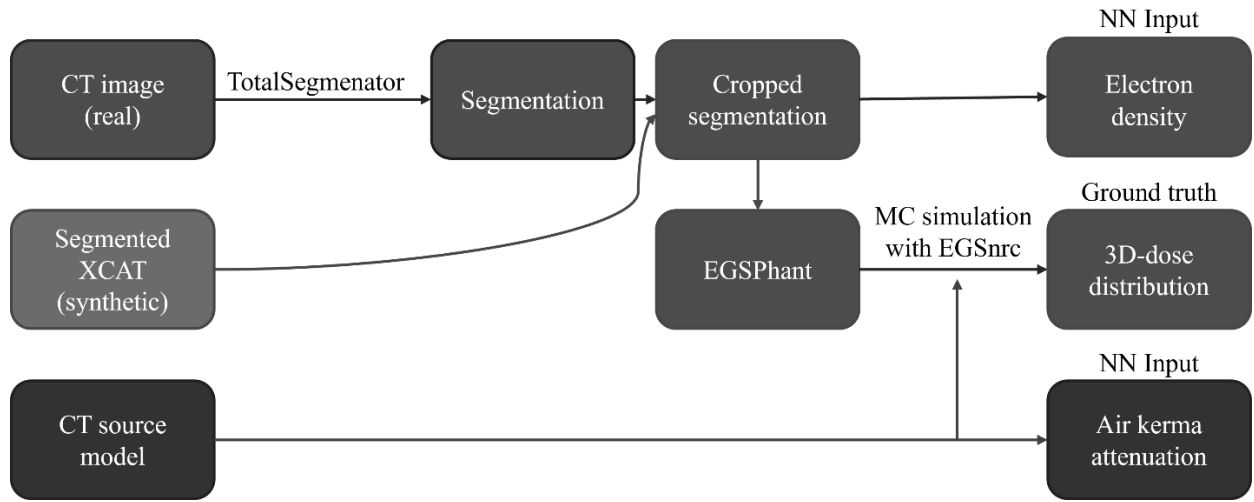

Figure 2. Data processing pipeline including segmentation and simulation of input and target data for training of the neural networks.

The neural network received two inputs as shown in Figure 1 and Figure 3, the first of which was the electron density distribution of the patient geometry. For this, the tissues in the segmented geometry were converted into electron density values according to ICRU Report 46 [19] of 1992, Appendix A, Table A1. The second input represented the radiation field properties, including photon fluence spectrum and beam profile. The data preprocessing is illustrated in Figure 2.

The air kerma attenuation of the CT radiation field in a water cylinder with a diameter of 32 cm (typical diameter of an body CTDI phantom) was determined for one angle and one z-position. For this calculation, a linear projection algorithm based on the software implemented by Lehner *et al* [20] was used. To reduce computation time, the resulting kerma value map was rotated and translated to mimic the CT source positions and their respective dose contribution.

The network had two output channels representing the 3D dose distribution map and an estimated variance. For optimization the dose map was compared to simulated data.

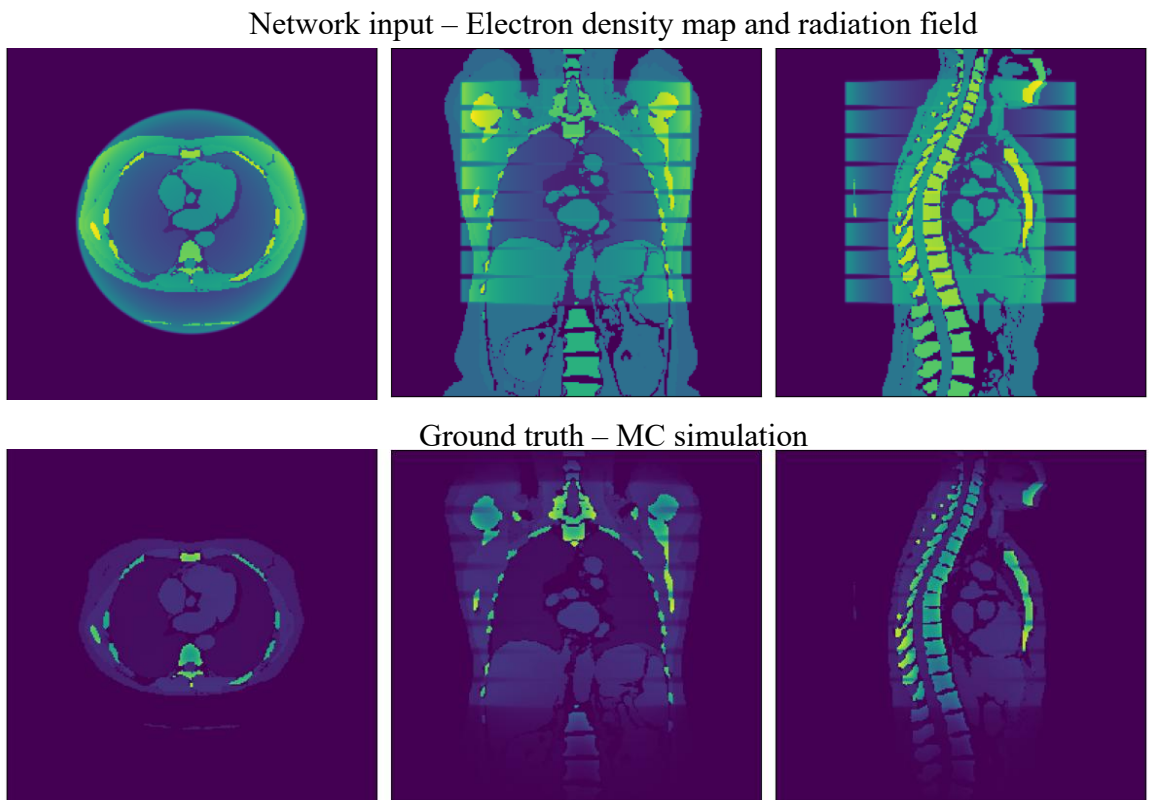

Figure 3. Example air kerma attenuation map for an axial CT scan, with an electron density distribution (top). Ground truth MC simulation (bottom).

In preparing for the training process, all data was normalized between 0.001 and 1 using min-max normalization. Due to computational constraints and to improve model generalization, the network was trained using patch-based processing instead of full-volume inputs. From each volume, random points in the non-air geometry were sampled and used as the middle points of 32×32×32 voxel patches.

*2.3.1. Network training*

To account for the spatially varying uncertainty in the voxel-wise dose estimation, a heteroscedastic loss function based on a negative Gaussian log-likelihood formula was used. The neural network was designed to calculate the predicted mean dose $\mu(x)$ and the corresponding logarithmic variance $\log \sigma^2(x)$ for each voxel $x$. The loss per voxel is given by:

$$\mathcal{L}(x) = \frac{1}{2}\exp(-\log \sigma^2(x))(y(x) - \mu(x))^2 + \frac{1}{2}\log \sigma^2(x). \qquad (1)$$

With this formulation, the model learned to express the uncertainty directly in terms of the predicted variance. The total loss for a single input patch was masked to include only voxels within the patient's body, as defined by a binary mask $M(x)$ computed from the segmentation map:

$$\mathcal{L}_{\text{total}} = \sum_x M(x)\mathcal{L}(x), \qquad (2)$$

where the summation runs over all voxels in a patch. Optimization was performed using the Adam optimizer with an initial learning rate of 1e-3 and batch size 32.

The model was trained for 1000 epochs. Checkpoints were saved every 100 epochs, and the best performing network was chosen from these ten networks based on organ dose prediction accuracy. The model was trained on an NVIDIA RTX A5000 GPU with an Intel Core i9-14900K CPU and 128 GB RAM. The system ran on Windows 11 Pro, utilizing PyTorch 2.3.1 with CUDA 12.1.

Four networks, named after their percentage of real training data, were trained on different datasets, realizing different ratios of real and synthetic patient data: 0 % NN: 0/100, 10 % NN: 10/90, 20 % NN: 20/80, 100 % NN: 100/0. The number of training data slightly increases from the 0 % NN to the 20 % NN. The datasets were split into 85 % training data and 15 % validation data for the training process. The test dataset was predefined and consisted exclusively of real patient data from 28 individuals. All of the networks were evaluated using the same test dataset.

*2.3.2. Performance evaluation and uncertainty estimation*

Organ doses (see Table 4) were derived from the dose distributions obtained from the NNs and the MC ground truth data by calculating the mean voxel dose for each organ.

To estimate network performance, the summed organ doses were compared between MC and NN for each patient geometry in the test dataset by calculating the relative deviation.

The uncertainty estimation was performed according to [21]. The voxel-wise dose estimation uncertainty $\sigma_{\text{tot}}(x)$ can be split into two factors: model uncertainty (epistemic) and data uncertainty (aleatoric). The deep ensemble method [21] was used to estimate the model uncertainty. As such, each network was trained five times with five different seeds. The mean value across all networks was taken as the dose estimation result

$$D(x) = \frac{1}{5}\sum_{i=1}^{5}\mu_i(x) \tag{3}$$

for each voxel $x$, and the variance

$$\sigma^2_{\text{epistemic}}(x) = \frac{1}{5}\sum_{i=1}^{5}(\mu_i(x) - D(x))^2 \tag{4}$$

was calculated and used as the model uncertainty.

The data uncertainty was addressed by training a neural network using a heteroscedastic loss function as described in Equation (1). To enable a robust estimate of the uncertainty, the uncertainty of the Monte Carlo data ($5.5\ \%, k = 1$) was enforced as a minimum for the aleatoric uncertainty $\sigma_i$ of each ensemble network via

$$\log \sigma_i^2(x) = \log(\tilde{\sigma}^2{}_i(x) + (\mu_i(x) * 0.055)^2) \tag{5}$$

where we recall that $\tilde{\sigma}_i$ is the second output channel of the network.

The resulting mean value over all ensemble members was used as aleatoric uncertainty:

$$\sigma^2_{\text{aleatoric}} = \frac{1}{5}\sum_{i=1}^{5} \sigma_i^2(x). \tag{6}$$

The total uncertainty

$$\sigma_{\text{tot}}(x) = \sqrt{\sigma^2_{\text{epistemic}}(x) + \sigma^2_{\text{aleatoric}}(x)} \tag{7}$$

is given as the square root of the summed variances.

For the organ doses, each voxel uncertainty was propagated to calculate the organ and summed organ dose uncertainties. For scalability reasons, the uncertainties were calculated under the assumption of full uncertainty correlation amongst the voxels. The variance of the mean can therefore be calculated by

$$\sigma^2_{\text{organ}} = \frac{1}{N_{\text{organ}}^2} \sum_{i=1}^{N_{\text{organ}}} \sum_{j=1}^{N_{\text{organ}}} \sigma_{\text{tot}}(i)\sigma_{\text{tot}}(j) = \frac{1}{N_{\text{organ}}^2} \left(\sum_{i=1}^{N_{\text{organ}}} \sigma_{\text{tot}}(i)\right)^2. \tag{8}$$

To allow performance comparison for each patient from the test dataset and for each model, a summed organ dose

$$D_{\text{sum}} = \sum_{i=1}^{\text{\# organs}} D_{\text{organ } i} \tag{9}$$

was calculated.

In the following chapters, the maximum uncertainty computed over the entire test data set is presented as the overall uncertainty measure for each neural network.

# 3. RESULTS

## 3.1. Uncertainties

Table 3 summarizes the aleatoric, epistemic, and total uncertainties associated with the summed organ dose estimation of the four neural network models.

The uncertainty shows a clear dependence on the fraction of real data included during training. The aleatoric uncertainty component remains constant at around 6 % for models trained predominantly on synthetic data, with a noticeable reduction to 3.6 % for the NNs trained exclusively on real patient data.

Table 3. Maximal values for uncertainty components and total uncertainties of summed organ doses for the four different neural networks over the test dataset. Uncertainties are reported as relative values, expressed in percent.

| Uncertainty component | 0 % NN | 10 % NN | 20 % NN | 100 % NN |
|---|---|---|---|---|
| $\sigma_{\text{aleatoric}}$ | 6.1 % | 6.2 % | 6.3 % | 3.6 %* |
| $\sigma_{\text{epistemic}}$ | 6.2 % | 3.9 % | 2.6 % | 3.5 % |
| $\sigma_{\text{tot}}$ | 8.9 % | 7.3 % | 6.8 % | 5.0 % |
| $u_{\text{tot}}$ ($k = 2$) | 17.8 % | 14.6 % | 13.6 % | 10.0 % |

* The reported uncertainty is lower than the voxel-level uncertainty of 5.5% because the uncertainty decreases when propagated over multiple voxels.

The epistemic component shows a stronger dependence on real data incorporation, decreasing from 6.2 % for the 0 % NN to 2.6 % for the 20 % NN and reflecting the reduction of model uncertainty upon introduction of a more diverse training dataset.

As a result, the total standard uncertainty $\sigma_{\text{tot}}(x)$ decreases from 8.9 % to 5 %, corresponding to expanded uncertainties $u_{\text{tot}}(x)$ of 17.8 % and 10.0 %, respectively.

## 3.2. Network performance

Figure 4 shows box plots of the distribution of the summed organ dose estimations for the ground truth and the four neural networks. Each box plot represents the interquartile range, with the median marked and individual patient predictions overlaid as black dots. The 0% NN substantially underestimated the dose compared to the ground truth, while the models trained with more than 10 % real data demonstrate better agreement. The 0 % NN exhibits a typical out-of-distribution behaviour [22]. The distribution for all networks was narrower compared to that of the ground truth values. The bottom panel shows the mean dose values relative to the mean ground truth for each model (dashed line), along with the mean associated uncertainties ($k = 1$). The 0 % NN model displays the greatest deviation and the highest uncertainty. For all other networks, the mean values agree with the ground truth within the uncertainty range.

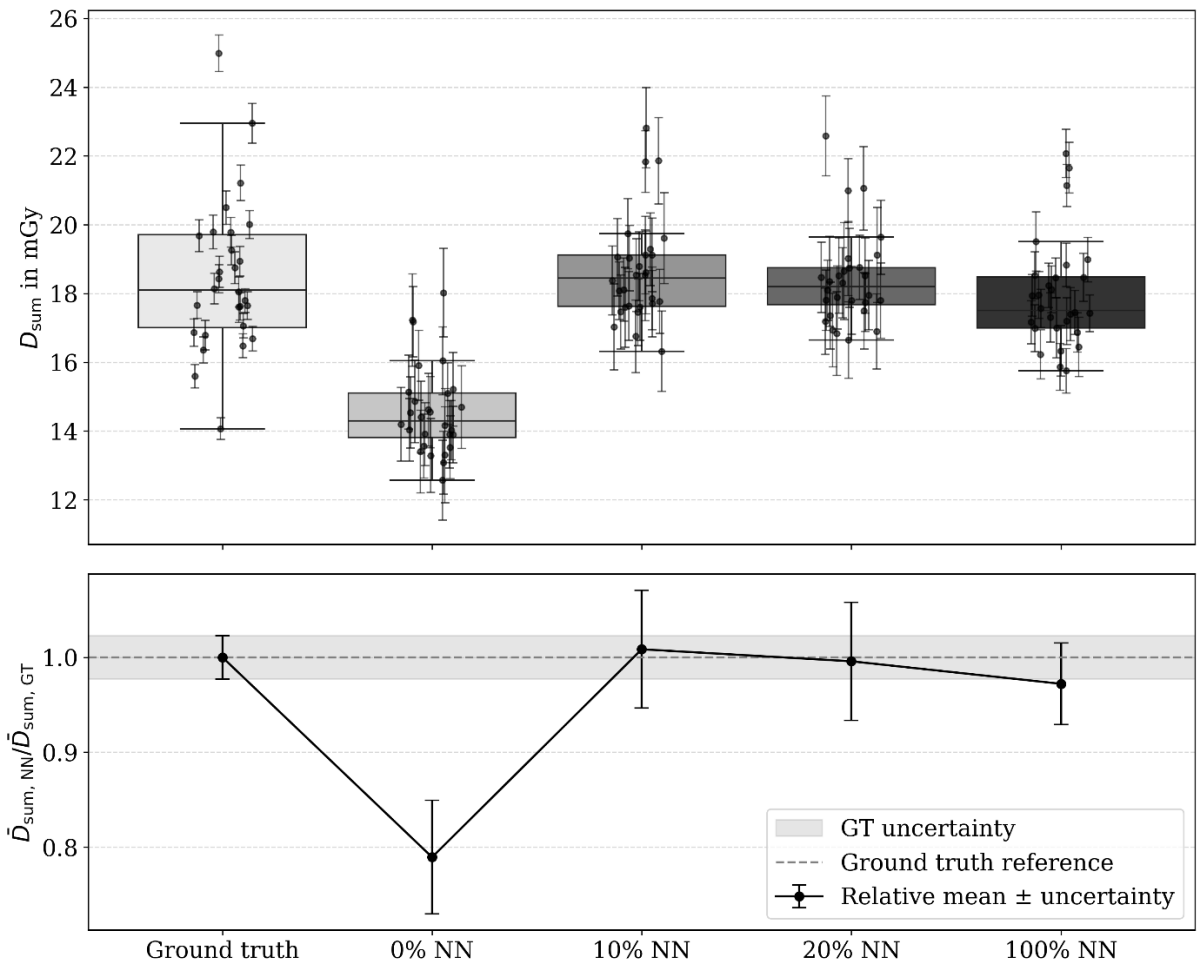

Figure 4. Model performance comparison for normalized summed organ doses across the test dataset via distribution of total doses, calculated for an assumed tube current of 200 mA (top); mean values relative to the ground truth MC data (bottom).

## 4. DISCUSSION

The results of this study demonstrate that the accuracy of neural network-based organ dose estimation is strongly influenced by the composition of the training dataset. The network trained exclusively on synthetic data showed poor agreement with the Monte Carlo reference and its total uncertainty was evaluated to be 17.8 % ($k = 2$). This result can be explained by out-of-distribution behaviour [22]. While the synthetic and real patient data appear similar overall, a notable difference emerges for the NNs. The uncertainty estimation of the 0 % NN was trained only on synthetic data and therefore underestimates the uncertainty on the real patient test dataset. Incorporating even a modest proportion of real patient data (10-20 %) led to a considerable increase in predictive accuracy, resulting in total uncertainties of less than 15 % ($k = 2$). An analysis of the uncertainty of the summed organ dose estimation reveals differing trends for the aleatoric and epistemic uncertainty components. The aleatoric data uncertainty remains nearly constant and only decreases in the pure real-data model, indicating that this component primarily reflects intrinsic data variability. In contrast, the epistemic uncertainty shows a pronounced drop from 6.2 % (0 % NN) to 2.6 % (20 % NN), confirming that even a small fraction of real patient data markedly improves model confidence. This result is similar to the one discussed by Khosravi *et al*, who saw the best performance when 10 % real data was incorporated into the training dataset [9].

An organ-level analysis, the results of which are presented in the Appendix (Table 4), demonstrated that predictive accuracy varied considerably across different anatomical regions. Small-volume organs (e.g., gallbladder, oesophagus) and organs located outside the primary irradiated field (e.g., stomach, colon, kidneys) exhibit higher relative deviations from MC dose estimates. These discrepancies are likely due to a combination of factors, including increased inter-patient anatomical variability, lower absolute dose levels, and reduced signal-to-noise ratios in the MC ground truth for such regions. In most cases, larger deviations were the result of an overestimation by the NNs.

Furthermore, performance differences across patient body habitus were evident. All networks showed a similar effective diameter dependency as seen in Figure 5.

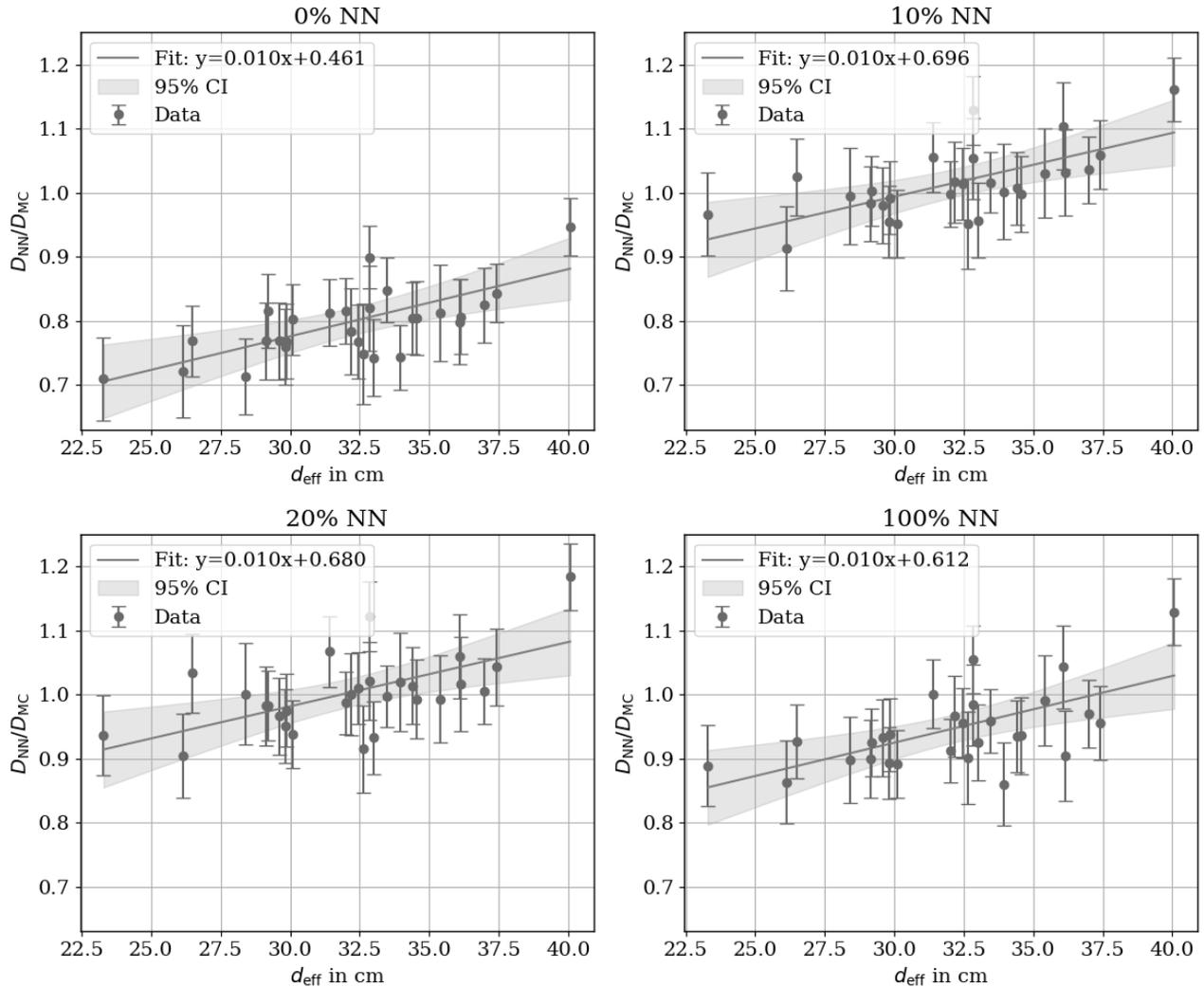

Figure 5. Ratio of NN-estimated summed dose to MC-based summed dose as a function of patient effective diameter. The 95 % confidence interval (CI) depicts the fit parameter uncertainties.

A key limitation here lies in the network's input design, in which the radiation field characterization was restricted to a fixed cylindrical input of 32 cm in diameter. For overweight patients, a substantial portion of the anatomy, and hence of potential dose-relevant structures, falls outside this input region. For very small patients, the radiation attenuation is not suitable. Addressing this limitation will require adapting the input geometry to be patient-specific, or incorporating variable field-of-view mechanisms that can better account for large body sizes.

Figure 6 shows an example of the estimated dose maps on a real patient anatomy over the four trained NNs. The out-of-field deviation is visible here as well. All NNs capture the general spatial structure of the MC reference, but small visible differences emerge between the different networks, especially in the peripheral areas. High-dose regions in bone show higher deviations for the 0 % NN, which is also reflected in the higher uncertainty in this area. Notably, the uncertainty maps closely mirror the spatial patterns seen in the deviation maps, with regions exhibiting larger deviations from the ground truth tending to also show higher uncertainty, especially in the areas outside the primary radiation field.

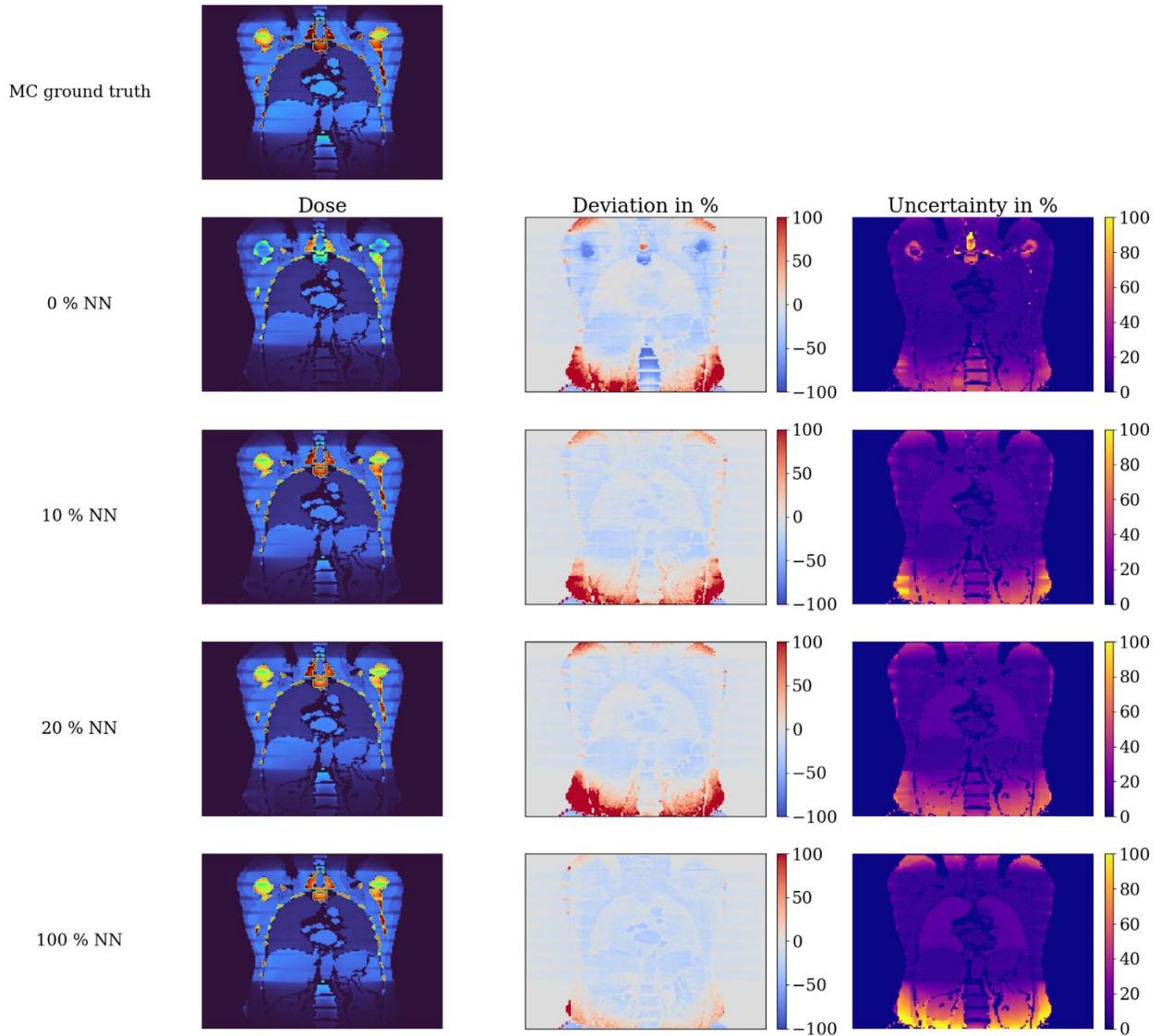

Figure 6. Example of a predicted dose map based on a real patient geometry (left). The corresponding relative deviations between NN and MC are shown in the middle. The right-hand column shows the uncertainties.

In summary, these findings suggest that mixed datasets incorporating even limited real patient data can significantly boost model performance, particularly when aiming for clinically acceptable accuracy across a broad patient population. However, future work should address persistent challenges in predicting doses for anatomically variable or low-dose regions. Targeted data augmentation, patient-aware input scaling, or modular architectures could help mitigate these limitations and improve robustness.

## 5. CONCLUSIONS

This study demonstrates that for our preprocessing approach and model design synthetic anatomical models are not sufficient on their own for training accurate deep learning models for personalized CT organ dose estimation due to out-of-distribution behaviour. A neural network trained exclusively on synthetic data exhibited limited accuracy, particularly for small organs and out-of-field structures and underestimates the uncertainty for real patient data. However, introducing even a small proportion of real patient data — as little as 10 % — led to significant improvements in both the statistical accuracy and spatial fidelity of dose predictions.

The controlled anatomical variability offered by synthetic datasets contributes to model generalization, while real patient data anchors the training to clinically realistic geometries and dose distributions. This hybrid training approach enhances robustness across diverse patient anatomies.

Overall, this work provides evidence that the integration of real and synthetic data is a viable path forward in personalized CT dosimetry. It reduces the need for large amounts of real patient data, which are often difficult to obtain, while still achieving a level of model performance suitable for adult patient dose assessment according to TRS-457 with a $k = 2$ uncertainty of less than 20 %.

## 6. APPENDIX

Table 4. Mean deviation in percent of organ doses between NN estimation and MC calculation

|  | 0 % NN | 10 % NN | 20 % NN | 100 % NN |
|---|---|---|---|---|
| Muscle | -11.1 | -4.1 | -4.6 | -4.1 |
| Lung | 25.8 | 16.0 | 17.7 | 16.0 |
| Fat | 5.7 | 6.6 | 4.6 | 6.6 |
| Blood | -1.3 | 5.6 | 1.3 | 5.6 |
| Heart | -10.52 | -1.7 | -4.4 | -1.7 |
| Kidney | 17.00 | 17.4 | 18.4 | 17.4 |
| Liver | -3.1 | 0.7 | 0.8 | 0.7 |
| Pancreas | -2.3 | 9.0 | 5.0 | 9.0 |
| Brain | 4.6 | 12.1 | 7.2 | 12.1 |
| Spleen | -0.9 | -0.7 | -3.9 | -0.7 |
| Red Marrow | -37.4 | -7.2 | -8.0 | -7.2 |
| Stomach | 844 | 755 | 800 | 755 |
| Gallbladder | 928 | 983 | 1074 | 983 |
| Adrenal Gland | 31.9 | 22.2 | 23.4 | 22.2 |
| Colon | 28.2 | 25.4 | 30 | 25.4 |
| Oesophagus | 2085 | 2089 | 2000 | 2089 |
| Part. effective dose* | 26.0 | 0.76 | 1.6 | 2.1 |

* The segmentation software did not accommodate all the organs necessary for effective dose calculation.

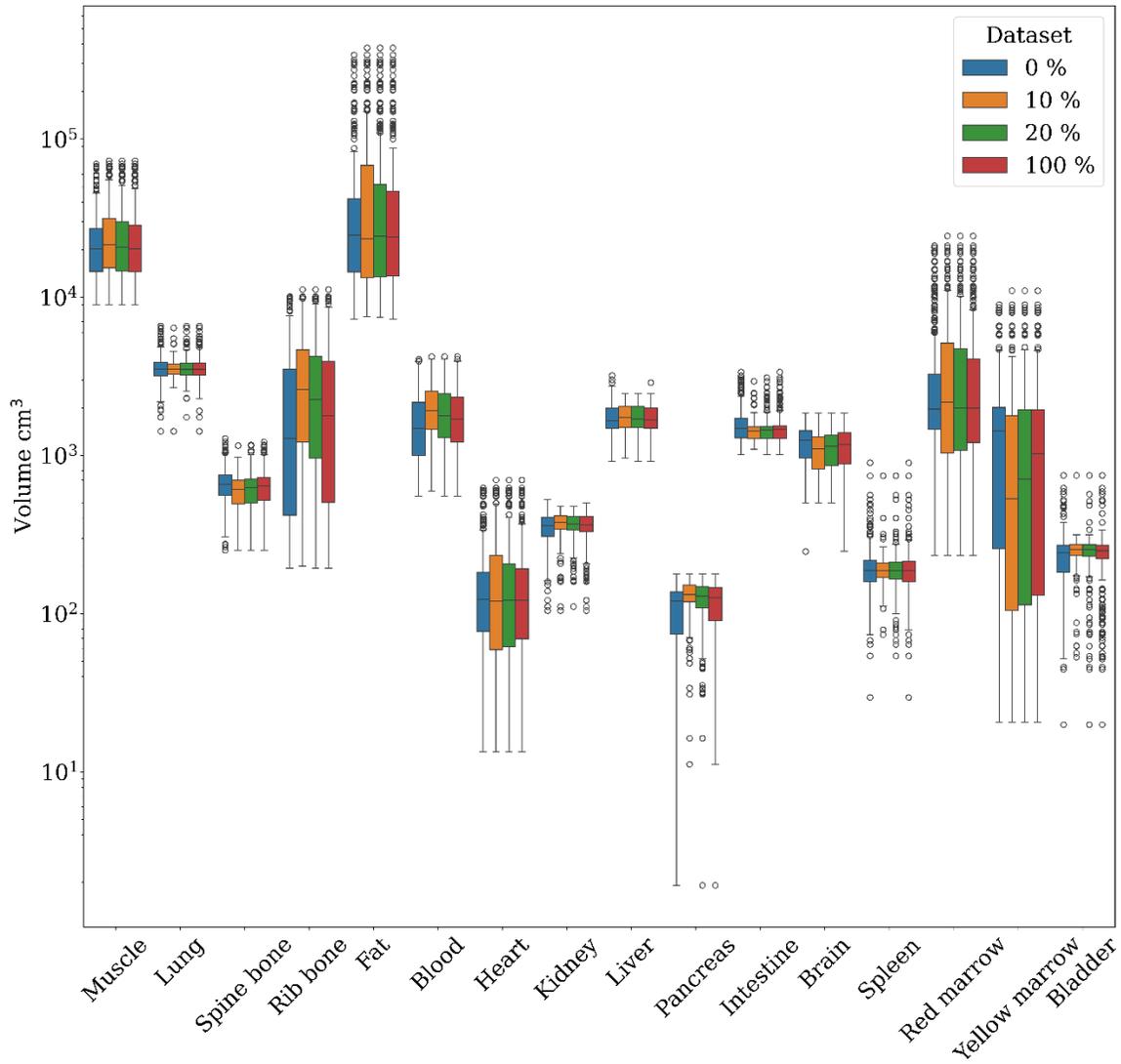

Figure 7. Comparison of organ volume distribution for the four different training datasets.

# References

[1] 'TRS 457 - Dosimetry in Diagnostic Radiology: An International Code of Practice', IAEA, Technical Report 457, 2007. Accessed: Apr. 30, 2025. [Online]. Available: https://www-pub.iaea.org/MTCD/Publications/PDF/TRS457_web.pdf

[2] J. Maier, L. Klein, E. Eulig, S. Sawall, and M. Kachelrieß, 'Real-time estimation of patient-specific dose distributions for medical CT using the deep dose estimation', *Med. Phys.*, vol. 49, no. 4, pp. 2259–2269, Feb. 2022, doi: 10.1002/mp.15488.

[3] E. Tzanis and J. Damilakis, 'A machine learning-based pipeline for multi-organ/tissue patient-specific radiation dosimetry in CT', *Eur. Radiol.*, vol. 35, no. 2, pp. 919–928, Aug. 2024, doi: 10.1007/s00330-024-11002-0.

[4] Y. Salimi, A. Akhavanallaf, Z. Mansouri, I. Shiri, and H. Zaidi, 'Real-time, acquisition parameter-free voxel-wise patient-specific Monte Carlo dose reconstruction in whole-body CT scanning using deep neural networks', *Eur. Radiol.*, vol. 33, pp. 9411–9424, June 2023, doi: 10.1007/s00330-023-09839-y.

[5] E. Sizikova *et al.*, 'Synthetic data in radiological imaging: current state and future outlook', *BJR|Artificial Intelligence*, vol. 1, Jan. 2024, doi: 10.1093/bjrai/ubae007.

[6] M. Frid-Adar, I. Diamant, E. Klang, M. Amitai, J. Goldberger, and H. Greenspan, 'GAN-based synthetic medical image augmentation for increased CNN performance in liver lesion classification', *Neurocomputing*, vol. 321, pp. 321–331, Dec. 2018, doi: 10.1016/j.neucom.2018.09.013.

[7] Y. Salimi, I. Shiri, Z. Mansouri, and H. Zaidi, 'Development and validation of fully automated robust deep learning models for multi-organ segmentation from whole-body CT images', *Physica Medica*, vol. 130, p. 104911, Feb. 2025, doi: 10.1016/j.ejmp.2025.104911.

[8] T. Russ *et al.*, 'Synthesis of CT images from digital body phantoms using CycleGAN', *International Journal of Computer Assisted Radiology and Surgery*, vol. 14, no. 10, pp. 1741–1750, Oct. 2019, doi: 10.1007/s11548-019-02042-9.

[9] B. Khosravi *et al.*, 'Synthetically enhanced: unveiling synthetic data's potential in medical imaging research', *eBioMedicine*, vol. 104, p. 105174, 2024, doi: https://doi.org/10.1016/j.ebiom.2024.105174.

[10] W. P. Segars, G. M. Sturgeon, S. Mendonca, J. Grimes, and B. M. W. Tsui, '4D XCAT phantom for multimodality imaging research.', *Med. Phys.*, vol. 37, no. 9, pp. 4902–4915, 2010, doi: 10.1118/1.3480985.

[11] C. H. Kim *et al.*, 'Adult Mesh-type Reference Computational Phantoms', *Annals of the ICRP*, vol. 49, no. 3, 2020, [Online]. Available: https://www.icrp.org/page.asp?id=575

[12] M.-L. Kuhlmann and S. Pojtinger, 'Generation of a representative synthetic phantom dataset for the training of neural networks in personalized CT dosimetry', in *Proceedings Virtual Imaging Trials in Medicine 2024*, Durham, NC: arXiv, May 2024. doi: 10.48550/ARXIV.2405.05359.

[13] J. Wassethal *et al.*, 'TotalSegmentator: Robust Segmentation of 104 Anatomic Structures in CT Images', *Radiol.: Artif. Intell.*, vol. 5, no. 5, Sept. 2023, doi: 10.1148/ryai.230024.

[14] I. Kawrakow, D. Rogers, E. Mainegra-Hing, F. Tessier, R. Townson, and B. Walters, *EGSnrc toolkit for Monte Carlo simulation of ionizing radiation transport*. (2000). C++,

Fortran. National Research Council of Canada. [Online]. Available: https://nrc-cnrc.github.io/EGSnrc/

[15] M.-L. Kuhlmann and S. Pojtinger, 'Implementation of a new EGSnrc particle source class for computed tomography: validation and uncertainty quantification', *PMB*, vol. 69, p. 095021, Apr. 2024, doi: 10.1088/1361-6560/ad3886.

[16] I. Sechopoulos *et al.*, 'RECORDS: improved Reporting of montE CarlO RaDiation transport Studies: Report of the AAPM Research Committee Task Group 268', *Medical Physics*, vol. 45, no. 1, Jan. 2018, doi: https://doi.org/10.1002/mp.12702.

[17] M. J. Berger *et al.*, 'XCOM-Photon Cross Sections Database, NIST Standard Reference Database 8 (XGAM) NIST, PML, Radiation Physics Devision'. Nov. 2010. doi: 10.18434/T48G6X.

[18] O. Ronnenberger, P. Fischer, and T. Brox, 'U-Net: Convolutional Networks for Biomedical Image Segmentation', *MICCAI*, vol. 9351, 2015, doi: https://doi.org/10.1007/978-3-319-24574-4_28.

[19] 'Photon, Electron, Proton and Neutron Interaction Data for Body Tissues', International Commission on Radiation Units and Measurements, ICRU Report 46, 1992.

[20] F. Lehner, P. Lombardo, S. Castillo, O. Hupe, and M. Magnor, 'RadField3D: a data generator and data format for deep learning in radiation-protection dosimetry for medical applications', *Journal of Radiological Protection*, vol. 45, no. 2, p. 021508, May 2025, doi: 10.1088/1361-6498/add53d.

[21] Lakshminarayanan, Balaji, Pritzel, Alexander, and Blundell, Charles, 'Simple and Scalable Predictive Uncertainty Estimation using Deep Ensembles', presented at the 31st Conference on Neural Information Processing Systems, Long Beach, CA, USA: arXiv, 2017. doi: 10.48550/ARXIV.1612.01474.

[22] D. Hendrycks and K. Gimpel, 'A Baseline for Detecting Misclassified and Out-of-Distribution Examples in Neural Networks', presented at the ICLR 2017, Toulon, France, 2016. doi: https://doi.org/10.48550/arXiv.1610.02136 Focus to learn more.

**DISCLOSURES OF CONFLICTS OF INTEREST**

The authors declare that their research was conducted without any commercial or financial relationships that could be construed as a potential conflict of interest.

**FUNDING**

No funding

**ACKNOWLEDGEMENTS**

The authors gratefully acknowledge the computing resources granted by PTB Sec. Q.45 High Performance Computing (HPC). The authors especially thank their colleagues Florian Burger and Gert Lindner of the HPC for their support during data production.

We gratefully acknowledge the Klinikum Braunschweig for generously providing the CT scan data used in this study.

**DATA AVAILABILITY STATEMENT**

The data supporting this study's findings are available from the corresponding author, M.-L. Kuhlmann upon request.